\documentclass{article}

\usepackage{microtype}
\usepackage{graphicx}
\usepackage{booktabs} 

\usepackage[]{changes}

\usepackage{enumitem}
\usepackage{wrapfig}
\usepackage{circledsteps}
\usepackage{xspace}
\usepackage{comment}
\usepackage{amsmath}
\usepackage[frozencache=true,cachedir=minted-cache]{minted}
\usepackage{algorithm}
\usepackage{algpseudocode}
\usepackage{subcaption}
\usepackage{caption}

\usepackage{mathtools}
\usepackage{multirow}
\usepackage{dsfont}
\usepackage[precision=2, unit=mm]{lengthconvert}

\usepackage{tikz}
\usetikzlibrary{patterns}
\usepackage{colortbl}

\definecolor{Gray}{gray}{0.9}
\definecolor{White}{gray}{1}
\definecolor{CommentColor}{HTML}{004564}

\usepackage{siunitx}

\usepackage{titlesec}

\titlespacing*{\section}{0pt}{*0.3}{*0.3}
\titlespacing*{\subsection}{0pt}{*0.2}{*0.2}
\titlespacing*{\subsubsection}{0pt}{*0.15}{*0.15}

\usepackage{hyperref}

\usepackage[accepted]{mlsys2024}

\mlsystitlerunning{\codename: Preserving Model Confidentiality during Graph Optimizations}

\newcommand{\codename}[0]{\textsc{Proteus}\xspace}
 
\usepackage{inconsolata}

\begin{document}

\twocolumn[
\mlsystitle{\codename: Preserving Model Confidentiality during Graph Optimizations}



\mlsyssetsymbol{equal}{*}

\begin{mlsysauthorlist}
\mlsysauthor{Yubo Gao}{uoft,vector,centml}
\mlsysauthor{Maryam Haghifam}{uoft}
\mlsysauthor{Christina Giannoula}{uoft,centml}
\mlsysauthor{Renbo Tu}{uoft}
\mlsysauthor{Gennady Pekhimenko}{uoft,vector,centml}
\mlsysauthor{Nandita Vijaykumar}{uoft,vector}

\end{mlsysauthorlist}

\mlsysaffiliation{uoft}{Department of Computer Science, University of Toronto}
\mlsysaffiliation{vector}{Vector Institute}
\mlsysaffiliation{centml}{CentML Inc}
\mlsyscorrespondingauthor{Yubo Gao}{ybgao@cs.toronto.edu}
%
\mlsyskeywords{Machine Learning, MLSys}

\vskip 0.3in


\begin{abstract}
\vspace{-15pt} 
Deep learning (DL) models have revolutionized numerous domains, yet optimizing them for computational efficiency remains a challenging endeavor. Development of new DL models typically involves two parties: the model developers and performance optimizers. The collaboration between the parties often necessitates the model developers exposing the model architecture and computational graph to the optimizers. However, this exposure is undesirable since the model architecture is an important intellectual property, and its innovations require significant investments and expertise. During the exchange, the model is also vulnerable to adversarial attacks via model stealing.

This paper presents \codename, a novel mechanism that enables model optimization by an independent party while preserving the confidentiality of the model architecture. 
\codename obfuscates the protected model by partitioning its computational graph into subgraphs and concealing each subgraph within a large pool of generated realistic subgraphs that cannot be easily distinguished from the original. We evaluate \codename on a range of DNNs, demonstrating its efficacy in preserving confidentiality without compromising performance optimization opportunities. \codename effectively hides the model as one alternative among up to $10^{32}$ possible model architectures, and is resilient against attacks with a learning-based adversary. We also demonstrate that heuristic
based and manual approaches are ineffective in identifying the protected model.
To our knowledge, \codename is the first work that tackles the challenge of model confidentiality during performance optimization. \codename will be open-sourced for direct use and experimentation, with easy integration with compilers such as ONNXRuntime. 
\vspace{-20pt}
\end{abstract}
]



\printAffiliationsAndNotice{}  
\newenvironment{ap}{\color{MidnightBlue}}{}

\section{Introduction}

Deep learning (DL) has emerged as a highly effective approach with a wide range of use cases. The remarkable performance achieved by DL models in domains like computer vision, natural language processing, and recommendation systems has immensely fueled their popularity. Models for ChatGPT, stable diffusion~\cite{diffusion}, and vision transformers~\cite{vit}, all demonstrate the potential of DL models in solving complex tasks. This has led to widespread interest in the generation of new models and DL innovations for novel and more powerful capabilities in both academia and industry.

A major challenge with DL models is the significant computational overhead for training and inference. DL models may have millions to hundreds of billions of parameters~\cite{gpt3}, requiring significant memory resources and compute. Thus, training DL models and deploying trained models for inference can be extremely expensive and time-consuming. This issue is expected to be exacerbated in the future, as model sizes continue to grow. 
For example, OpenAI reports a daily cost of $\$700$K to run ChatGPT~\cite{chatgpt_700k}.

Performance optimizations using ML compilers have thus become crucial to efficient training and inference to reduce latency, computational expenses, and energy consumption. Recently developed optimizing compilers/tools include TVM~\cite{tvm}, TASO~\cite{taso}, ONNXRuntime~\cite{onnxruntime}, and Hidet~\cite{ding2023hidet} and is an active area of research and development. 
Existing tools are already proven to be highly effective in generating significant speedups and are thus widely used. For example, TVM can provide up to $3.8\times$ speedup on model inference~\cite{tvm}. 

Model optimization and model creation/development are typically not done at the same time by \emph{the same party}. Model development and optimization each requires different expertise and domain knowledge. For example, while model developers are good at designing neural network architectures, they often do not possess the domain skills necessary for performance optimization. This has led to the emergence of companies offering model optimization as a service such as OctoML~\cite{octoml} and MosaicML~\cite{mosaicml} to fill in the gap. 

Several existing compilers \cite{tvm,onnxruntime,ding2023hidet,TensorRT} can be used to \emph{automatically} optimize the model, potentially enabling model developers to directly produce performant implementations without a second party for optimization.
However, \emph{solely} relying on automatic compilers has limitations and does not eliminate the need for additional optimization expertise. First, effectively optimizing tensor computations is still challenging even when using an automatic compiler and often requires significant domain expertise and intervention. For example, correctly configuring the tensor compiler (e.g., selecting search space, using the correct floating point precision), adding previously-unsupported operators~\cite{tvm_addop}, or implementing scheduling templates for novel operators~\cite{ding2023hidet} requires systems expertise. Second, they are less effective at optimizing for proprietary hardware or at leveraging hardware features that are not fully exposed by hardware vendors~\cite{Ansor}, and thus may require specialized expertise from hardware vendors about their hardware. For example, the optimizations/libraries specific to Google's TPUs in XLA and NVIDIA GPUs are closed source and require support from the hardware vendors when the provided tools are not effective~\cite{tfxla_issues}. Third, the developers of novel optimizing tools may not provide open-source implementations or the entire toolset for automatic use due to proprietary optimizations or the need for manual intervention. For example, OctoML~\cite{octoml} applies proprietary optimizations manually for customers\cite{octoml_wonbo}, a process which requires manual insight. 

The necessity of two parties to effectively develop and optimize new DL models leads to an important novel challenge: ensuring confidentiality of the DL architecture. 
Highly effective performance optimizations include graph-level transformations which involve optimizing the computational graph (i.e. the graph of operators). Possible transformations include techniques such as operator fusion, constant folding, and functional approximations~\cite{ort_graphopts}. Graph level performance optimizations typically require providing the optimizing party direct access to the entire computational graph of the model. However, the model architecture itself is expensive intellectual property to the model developers, as innovating novel DL architectures requires domain experts and extensive resources for neural architecture search and training. For example, NASNet~\cite{nasnet} is discovered through thousands of GPU days spent on neural architecture search, and a single training of GPT-3 costs $\$4.6$M\cite{gpt3}. Additionally, exposing the model architecture exacerbates the threat of adversarial attacks~\cite{adversary} by model stealing approaches that can then be used to perform gradient-based adversarial attacks~\cite{gradadversary}. 

In this work, we present \codename, an obfuscation mechanism that aims to preserve the confidentiality of the protected model during graph optimizations.
\codename effectively enables an independent party with a proprietary optimization tool such as a machine learning compiler to optimize a novel model architecture with no direct knowledge of the original model architecture. 
\codename is largely agnostic to the optimizations themselves and can be generally used by any optimization tool. 

\begin{figure*}[t]
    \centering
    \includegraphics[width=0.75\textwidth]{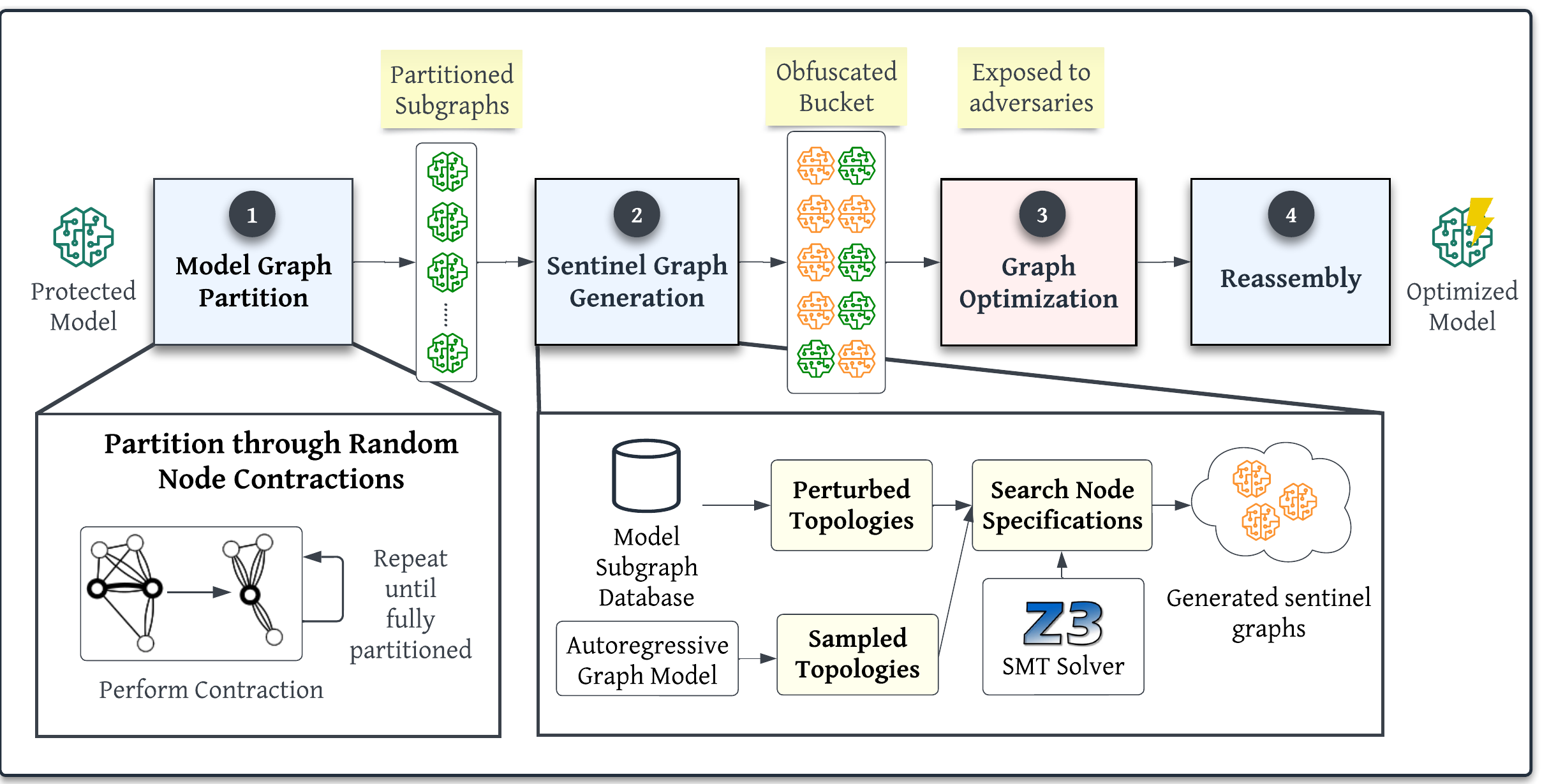}
    \vspace{-6pt}
    \caption{System Overview of \codename}
    \vspace{4pt}
    \label{fig:overview}
\end{figure*}

The key idea behind \codename is twofold. First, we propose to generate \emph{sentinel} graphs, which are artificially generated graphs that resemble real world DL computational graphs. These sentinel graphs are provided alongside the original graph to the optimizing party such that the optimizing party cannot distinguish which graph is the \emph{protected} graph. This approach alone, however, still involves providing the optimizing party with the protected graph in its entirety.

To address this challenge, our second idea leverages \emph{graph partitioning}. Graph partitioning first partitions the protected graph into smaller subgraphs. We then generate sentinel subgraphs for each protected subgraph. The optimizing party is now provided with a bucket of sentinel subgraphs and protected subgraphs for optimization that are indistinguishable from each other. This approach requires that the adversary would have to correctly identify \emph{every} protected subgraph to recover the protected model. At the same time, the optimizing party can optimize each of the subgraphs flexibly. The model owners can then trivially reconstruct the original model from the optimized subgraphs.

We demonstrate that with sentinel generation and graph partitioning, it would be infeasibly expensive to correctly identify the original protected graph. We illustrate the major steps of \codename in Figure \ref{fig:overview}. At a high level, \codename accepts a model to be optimized by the optimizer party. The obfuscation mechanism converts the graph that needs to be optimized (hereafter referred to as the ``protected'' graph) into a set of subgraphs that contain both parts of the protected graph (the ``protected subgraphs'') as well as artificially generated subgraphs (the ``sentinel subgraphs''). The optimizer-party then performs optimizations on the collection of obfuscated subgraphs (indistinguishably includes both the original and artificial subgraphs). The optimized subgraphs are returned to the model owner who can trivially assemble the optimized protected graph. 

There are several major criteria that need to be met in designing \codename. First, it is crucial that the generated sentinel subgraphs are difficult to differentiate from the protected subgraphs, while still ensuring that no general information regarding the protected graph can be inferred. Second, it is important to perform graph partitioning such that optimization opportunities are not lost, while generating enough subgraphs to sufficiently obfuscate the protected graph. We discuss how \codename addresses these challenges in Section 4 and develop a mechanism that allows trading off obfuscation quality for less optimization overhead. 

We evaluate \codename using a range of common image and language models to evaluate the effectiveness of \codename's obfuscation mechanism, we devise an adversarial attack using a learned classifier model to distinguish between real subgraphs and the generated sentinels. We demonstrate that such an attack is unable to distinguish the real protected model from a pool of $10^7$ to $10^{32}$ potential model architectures, making recovery computationally infeasible using this method. Across all the evaluated models, \codename retains the ability of the optimizer to provide significant speedups via graph-level optimization, with an average speedup within $10\%$ of the maximum attainable by the optimizer. We demonstrate \codename's overall use and effectiveness using two case studies.{, which show that \codename remains within $1\%$ and $10\%$ of the speedup of the optimizer. In both cases, our method requires a learning-based adversary to evaluate more than $10^{23}$ models in which the original model is hidden.} We will provide an open-source implementation of the tool that can be directly used with modern optimizing compilers and can serve as basis for future research on this topic.

To summarize, this work makes the following contributions:
\begin{enumerate}[label=(\alph*), leftmargin=*,noitemsep,topsep=0pt]
    \item We motivate the need for a mechanism that effectively decouples model innovation and model optimization by preserving the confidentiality of the model architecture. Such a mechanism would flexibly enable model development and performance optimization to be performed by independent parties, without the optimization party having full knowledge of the confidential model architecture. 

    \item We propose \codename, the first mechanism to tackle this challenge of preserving confidentiality of any arbitrary DL model during performance optimization. \codename partitions the protected DL model into subgraphs, and hides them within sentinel graphs. \codename also 
    effectively preserves the efficacy of various graph-level optimizations performed by optimizers.
    
    \item We propose a novel subgraph generation tool that is able to produce realistic artificial subgraphs to obfuscate the original subgraph. To demonstrate the robustness of our approach, we devise a learning-based attack attempting to identify the sentinels and demonstrate that it is ultimately ineffective in recovering the protected DL model. We also demonstrate that heuristic based and manual approaches are ineffective in identifying the protected model.
\end{enumerate}
\section{Background and Related Work}
\vspace{-2pt}
 \subsection{Graph-Level Optimizations for DL Models} 
 \vspace{-2pt}
 
Deep learning (DL) compilers~\cite{tvm,taso,ding2023hidet,sabne2020xla,glow,ngraph,sparsetir} provide graph-level and operator-level optimizations for DL models to accelerate their deployment and system performance. To apply these optimizations, the DL compiler operates on the graph representation of the model, i.e., the architectural structure of the model that determines how the layers and connections are organized to process input data and generate the desired outputs. A DL model is typically expressed as a directed acyclic graph (DAG), hereafter named as \emph{computational graph}, in which the nodes represent the DL operators (e.g., convolution, pooling, activation) and the edges represent the dependencies between the nodes, i.e., the tensors given as inputs/outputs to the operators~\cite{bengio2009learning}. DL compilers first apply graph-level optimizations on the computational graphs and then perform operator-level optimizations. The graph-level optimizations are rule-based transformations: they simplify executed computations, and they are either manually designed or automatically generated  using heuristics algorithms. For instance, TVM~\cite{tvm}, TensorRT~\cite{TensorRT}, and ONNXRuntime~\cite{onnxruntime} integrate generic rule-based transformations that assist in finding and applying optimizations. ONNXRuntime supports specific graph-level optimizations, such as Identity Elimination and Reshape Fusion. TASO~\cite{taso}, on the other hand, \emph{automatically} generates rule transformations for a given set of operators and verifies their correctness. Recently, companies (such as OctoML and MosaicML~\cite{octoml,mosaicml}) have emerged that provide such optimizations as a service, i.e., developing DL compilers, tools, and services that accelerate the performance of emerging DL models. 

\vspace{-4pt}
\subsection{Data Privacy Solutions}
\vspace{-2pt}
\textbf{Differential Privacy.} Differential Privacy (DP)~\cite{dwork2006differential,dwork2014algorithmic} provides privacy-preserving data analysis and learning, i.e., extracting useful statistics and information from a dataset, while protecting the identity and privacy of individuals in the dataset. Typical DP methods, e.g., Laplace mechanism~\cite{laplace}, inject controlled noise or randomness to the data or statistical analysis process to protect the privacy of individuals, while preserving the overall utility of the data. DP approaches~\cite{dpsgd} have been proposed to protect \emph{user data} that is used to train models. However, it is unclear how these approaches can protect confidentiality of DL model architecture as adding noise to it damages functional correctness of the model. 

\textbf{Homomorphic Encryption.} Homomorphic Encryption (HE)~\cite{gentry2009fully} allows computations to be performed on encrypted data without the need for decryption. Homomorphism-based transformations are structure-preserving transformations, i.e., HE-based schemes preserve the additive and multiplicative structures of the data. 
Therefore, even though HE-based methods encrypt the parameters \cite{gilad2016cryptonets,hesamifard2017cryptodl} of DL models (e.g., weights of matrices, tensor values), they do not encrypt operators and topology of the model. During performance optimization, HE-based methods can ensure the privacy of model parameters, but they cannot protect the model's structure. 

\subsection{Model Stealing Attacks} 

Model stealing consists of creating a functionally-equivalent model and carrying out gradient-based adversarial attacks on the model. 
Prior works~\cite{oh2019towards,papernot2017practical,tramer2016stealing} design algorithmic-level analysis to create functionally-equivalent models. 
The initial phase of important adversarial attacks involves the extraction of the network architecture (topology) of the DL model: given the topology (network architecture) of a DL model is known, stealing attacks can infer the values of model parameters, hyper-parameters, and even training data~\cite{tramer2016stealing,wang2018stealing}. 
There are also several model topology extraction attacks, including DeepSniffer~\cite{deepsniffer} and ReverseCNN~\cite{hua2018reverse}, which extract the model architecture by leveraging architectural hints.

\section{Our Proposal: \codename}

\subsection{Threat Model}\label{sec:threat_model}
\codename aims to protect the model architecture from exposure to third parties, where the risk of model exposure is incurred when the model leaves the model owner and is:
\begin{enumerate}[leftmargin=*,noitemsep,topsep=0pt]
    \item intercepted by a third party in transit from the model owner to the optimizer through conventional wiretapping techniques, or
    \item leaked by the optimization party to a malicious third-party. This includes the possibility where the optimizer party is also the party performing the attacks.
\end{enumerate}

At the same time, implementations of the optimizing compiler remain as crucial intellectual properties of the optimization service. Release of the optimizing compiler to the model owner incurs the risk of software piracy.

\subsection{Goals}\label{sec:problem_formulation}
Our goal in this work is to effectively decouple model innovation/development and performance optimization by enabling performance optimization \emph{while protecting the confidentiality of the model architecture.} Ensuring privacy of the model architecture enables an independent party/service to optimize DL models. 
Specifically, we aim to achieve the following goals with our proposed mechanism:
\begin{enumerate}[leftmargin=*,noitemsep,topsep=0pt]
    \item \textbf{Model Confidentiality.} Given a mechanism that \emph{obfuscates} the graph to prevent the retrieval of the original architecture: an adversary with access to both the obfuscating algorithm used and the obfuscated graphs produced by the confidentiality mechanism should not be feasibly able to retrieve the initially protected graph.
    \item \textbf{Agnosticity and Independence of Performance Optimizations.} An effective confidentiality mechanism should not constrain the optimizations that can be performed on the protected graph. This additionally enables the potential for preserving the confidentiality of the model optimizations and the compilers themselves.
    \item \textbf{High Performance Efficiency.}
    Since the key goal of the optimizer is to improve the runtime performance of DL models, the confidentiality mechanism needs to preserve the performance benefits and ensure similar speedups as that achieved without the obfuscation mechanism. 
    
    \item \textbf{Low Compilation Overhead.} The confidentiality mechanism should not cause a significant compilation overhead to the optimizer party or make the optimization process more challenging. In this work, we focus our efforts on making the overhead of confidential optimization by machine learning compiler feasible.
\end{enumerate}

\subsection{Overview} In this work, we propose \codename, the first obfuscation mechanism for DNN computational graphs for performance optimization. \codename involves optimization in three independent steps. First, the \emph{obfuscation} step where the original computation graph is ``obfuscated" such that an adversary cannot feasibly identify the original model, thus providing confidentiality. Second, the \emph{optimization} step is carried out flexibly and independently by the optimizer party on the obfuscated computational graph, providing performance speedups. Finally, the \emph{de-obfuscation} step where the original model is retrieved by the model owner in its optimized form. 

\subsection{Obfuscation: Key Ideas}
The key ideas behind \codename are twofold: \emph{sentinel generation} and \emph{graph partitioning}. We detail each below. 

\subsubsection{Sentinel Generation.} We propose to obfuscate the original graph by hiding the protected DL graph among a set of \emph{sentinel} graphs, i.e., artificially generated realistic graphs. The idea behind this approach is that an adversary will be unable to distinguish the real graph from the sentinel models. As the set of sentinel graphs grows larger, it is more challenging to identify the protected graph. This approach enables the optimizing party to optimize the obfuscated graph -- by essentially optimizing all the sentinel graphs. 

However, directly applying this approach has important limitations. First, the original protected graph is still directly exposed in its entirety. Second, this approach adds significant overhead to the optimizing party as all the sentinel graphs need to be optimized. Generating $k$ sentinel graphs requires the both optimizer and the adversary to carry out $\mathcal{O}(k)$ work, either to perform optimizations or to attempt recovery of the protected model. To address these limitations, we first perform \emph{graph partitioning} as described below.

\subsubsection{Graph Partitioning.} 
We observe that most graph-level transformations performed by tensor compilers are local: graph-level substitutions performed by compilers operate on an operator and its neighboring nodes. We leverage this observation to partition the computational graph into smaller subgraphs. These subgraphs are independently optimized and then reassembled to generate the entire optimized graph. We evaluate the implications on performance speedup in Section \ref{sec:eval-performance} and demonstrate that this only incurs small losses in performance speedups from optimizations. 
With graph partitioning, the sentinel graphs are now generated for subgraphs rather than the entire graph. 
Thus, our solution can be broken down into two steps: (i) we partition the protected model into smaller subgraphs and (ii) hide the protected subgraphs within a set of $k$ sentinel subgraphs.

The use of sentinel subgraphs makes the recovery of the original model significantly more challenging for the adversary because every subgraph has to be correctly classified and identified to reconstruct the original protected graph. Thus, we can use fewer sentinel graphs while still making it infeasible to recover the original model (we quantitatively demonstrate this in Section \ref{sec:design}). At the same time, the model architecture in its entirety is never exposed to the optimizer. 

\subsubsection{Design Challenges}
There are two major challenges in effectively obfuscating the protected graph as described above: 

\textit{(i) Effective partitioning strategy:} The number of subgraphs in the graph plays a key role in our ability to retain the optimization performance benefits. If the subgraphs are too small or if the partitioning eliminates optimization opportunities, the optimization schemes may be rendered less effective. At the same time, increasing the number of subgraphs obfuscates the graph more effectively and would thus require fewer sentinel graphs. 

\textit{(ii) Generating sentinel graphs:} It is crucial to generate sentinel subgraphs that are difficult to identify as \emph{artificial}. Thus, they must be \emph{realistic}, syntactically accurate, and resemble real world subgraphs in terms of operations, topology, etc. In other words, the sentinels cannot be arbitrarily generated. 
We next describe \codename's detailed design where we aim to address the above challenges.
 
\section{\codename: Detailed Design}\label{sec:design}
We describe the three key steps of \codename in more detail. Section \ref{sec:detailed_obfuscation} describes the obfuscation mechanism of \codename. Section \ref{sec:detailed_optim} describes the optimization step needed to be performed by the optimizer party on the obfuscated subgraphs produced by \codename. Section \ref{sec:detailed_deobfs} describes the de-obfuscation step performed by \codename to retrieve the original model in its optimized form, returning it to the model owner.

\subsection{Obfuscation}\label{sec:detailed_obfuscation}
\codename is an effective obfuscation mechanism that preserves confidentiality in DL models consisting of two major steps: (i) \emph{graph partitioning} splits the protected model into smaller subgraphs, and (ii) \emph{sentinel graph generation} hides the protected subgraphs within a set of $k$ sentinel graphs. Specifically, given an arbitrary DL model, \codename generates $n$ smaller subgraphs and hides each subgraph within a set of $k$ sentinel (artificially generated) subgraphs. This way \codename hides the given protected DL model within a set of $\mathcal{O}((k+1)^{n})$ possible computational graphs. 

\subsubsection{Graph Partitioning} \label{sec:detailed_partitioning}
\codename splits the protected model into $n$ subgraphs of similar sizes. Our key goal is to generate many subgraphs, such that the adversary cannot feasibly identify the original model, while at the same time not affecting the graph-level optimizations performed by the optimizer party. Note that with \codename the optimizer applies graph-level transformations at each subgraph individually, and cannot perform optimizations that span across multiple subgraphs. However, given that we do not have any information in advance on which graph-level transformations will be performed by the optimizer, we employ a randomized graph partitioning algorithm to split the computational graph to $n$ subgraphs.

We develop a graph partitioning algorithmic scheme inspired by the Karger-Stein (K-S) algorithm~\cite{karger}. K-S is a randomized algorithm that solves the minimum-cut problem on a graph. At each step, it selects a random edge from the graph and merges the two nodes connected by the selected edge into a single node. This step is called ``edge contraction", and it is iteratively repeated until $n$ nodes remain in the graph. When the algorithm terminates, each of the $n$ remaining nodes represents a subgraph of the initial graph.

However, since K-S algorithm is randomized, the resulting $n$ subgraphs may significantly vary in size. Creating subgraphs with high disparity in their sizes brings two key issues. First, very large subgraphs may cause confidentiality issues, since they can potentially reveal many useful information to the adversary related to the initial protected graph. Second, small subgraphs might cause performance issues, since optimizers cannot perform very efficient graph-level transformations on small graphs.
Therefore, we enhance the K-S algorithm to create $n$ subgraphs of almost equal sizes. Specifically, we perform multiple iterations of the K-S algorithm and at each iteration we evaluate the standard deviation of the sizes of the subgraphs created. Then, we keep the graph partitioning scheme that minimizes the disparity in the sizes of the subgraphs, to provide a more balanced and less informative graph partitioning.

\subsubsection{Sentinel Graph Generation}\label{sec:detailed_sentinel_gen}
In the sentinel graph generation step, \codename hides the protected subgraphs within a set of $k$ sentinel subgraphs. The generated sentinels must be syntactically correct to avoid immediate detection. Moreover, the sentinel graphs should resemble real world ones, so that the adversary cannot differentiate between real subgraphs (extracted from the initial protected model) and the sentinel subgraphs (artificially generated using \codename).

In this step, we generate $k$ sentinel graphs for each of the $n$ subgraphs. We denote the $n$ subgraphs extracted from the original protected model with ${G_1,\ldots,G_n}$. For each subgraph $G_i$, we generate $k$ sentinel graphs denoted with $G_{i}^{(1)}, \ldots, G_{i}^{(k)}$. In other words, \codename creates a bucket of $k+1$ subgraphs, one of them is the real subgraph $G_i$, and the remaining $k$ subgraphs are sentinel (artificially generated) subgraphs. Therefore, the total search space for subgraphs has a size of approximately $\mathcal{O}((k+1)^n)$.

The $k$ generated sentinel subgraphs $G_{i}^{(1)}, \ldots, G_{i}^{(k)}$ should resemble the original real subgraph $G_i$, such that the adversary should not be able to differentiate among them. We can categorize two types of metrics, which an adversary could use to identify the real subgraph $G_i$:
\begin{enumerate}[label=(\alph*), leftmargin=*,noitemsep,topsep=0pt]
    \item \textbf{Topological Information.} 
    A subgraph can be identified by the topological connections of its nodes. Specifically, computational DL graphs are acyclic graphs that describe the dataflow of DL operators, and their nodes typically have a small number of incoming edges.
    \item \textbf{Operator Information.} 
    A subgraph can be identified by the patterns of computations performed at its nodes. 
    In a computation DL graph, nodes represent the operation applied to a tensor. Therefore, real subgraphs follow similar computational patterns: e.g., a convolution operator is commonly followed by an activation.
\end{enumerate}

\begin{figure}
    \centering
    \includegraphics[width=0.4\textwidth]{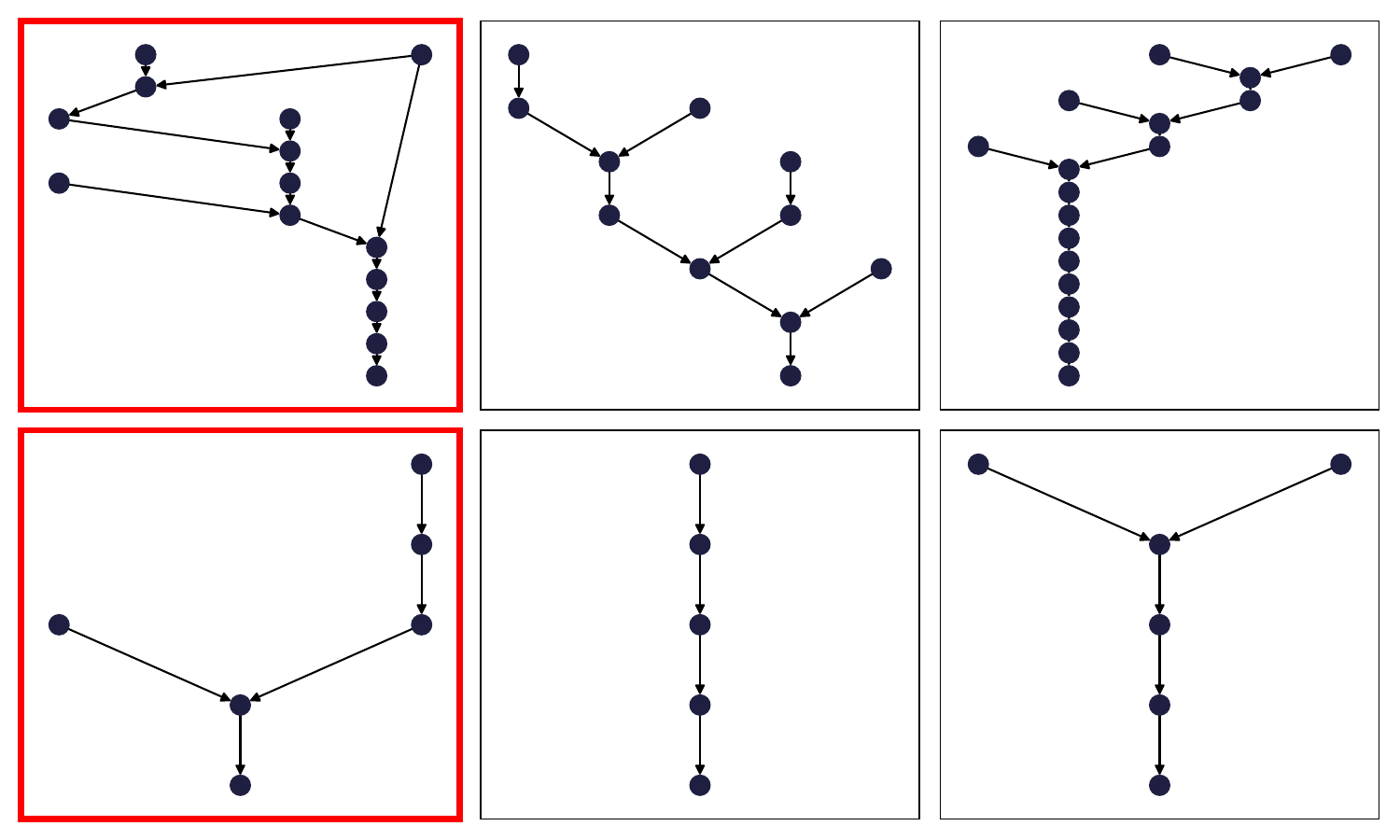}
    \caption{Examples of topologies sampled by \codename \\ {\small (red: the original topologies)}}
    \label{fig:generated_topology}
    \vspace{-15pt}
\end{figure}

Based on the two types of metrics listed above, \codename generates sentinel graphs in two stages. First, in the \emph{topology selection} stage, we generate the connections between the nodes in the computation graph. Second, we fill in operators into the graph in the \emph{operator population} stage.

\noindent \textbf{Topology Selection.} We first generate the topologies of the sentinel graphs. The topology refers to the ``shape'' of the computation graph, i.e., how the nodes are connected. The topology selection process for sentinel subgraphs has two major steps: graph generation and sampling. 

First, during the \emph{graph generation step}, we generate a pool of realistic graph topologies with GraphRNN~\cite{you2018graphrnn}, an autoregressive graph generation model. However, a key limitation of this GraphRNN-based approach is that it will generate \emph{undirected} graphs, while DL graphs are directed. To resolve this limitation, we transform the undirected graphs to \emph{directed} graphs using the algorithm shown in Algorithm \ref{alg:induce_orientation} in appendix section \ref{sec:appendix_induce_orientation}, which traverses the undirected graph and assigns a direction to each edge, resulting in a DAG. 

Next, we \emph{sample} sentinel topologies from the previous step that are similar to the provided real subgraph. Specifically, we approximate the similarity measure by comparing various graph- and node-level statistics of the sentinel subgraphs with that observed of real-world subgraphs. The evaluated similarity metrics are the following: average degree, clustering coefficient, diameter, and graph size. 

In algorithm \ref{alg:sample_topologies}, the \textsc{SampleTopologies} function takes as input a protected subgraph $G_i$ and generates a set of similar graph topologies that are statistically indistinguishable from $G_i$. This is achieved by ensuring that the graph statistics form a uniform distribution around the protected subgraph $G_i$, effectively adding random noise resulting in uncertainty. In other words, by observing the distribution of these subgraph statistics, each subgraph would have an equal chance of being the protected subgraph. Here, we extract a set of GraphRNN-generated graph topologies, denoted as $\mathcal{D}$, and control the range of the uniform distribution with $\beta$.

In algorithm \ref{alg:sample_topologies}, we establish bounds for uniform distribution in lines $2-8$, sample from this uniform distribution in lines $15-17$. Notably, if we sample graphs from $\mathcal{D}$ uniformly at random, the resulting distribution would follow that of $\mathcal{D}$ rather than being uniform. To tackle this, we employ \emph{importance sampling} which applies a weight of $1/p$ to each sample where $p$ describes the density 
 of the topology under $\mathcal{D}$. This ``corrects'' the uneven densities under $\mathcal{D}$ and makes the resulting samples uniform. 

\newcommand{\algcmt}[1]{{\emph{\color{CommentColor}#1}}}

{
\begin{algorithm}[t]
    \caption{Sampling similar topologies}
    \label{alg:sample_topologies}
    \begin{algorithmic}[1]
        \Function{SampleTopologies}{$G, \mathcal{D}, \beta$}
        \State\algcmt{Estimate the density $p$ from the GraphRNN graphs}
        \State $p(\mathbf{x}) \leftarrow \textsc{EstimateDensity}(\mathcal{D})$ 
        \State\algcmt{Sample the random position of $G$}
        \State Sample $\alpha\sim \operatorname{Unif}([0,\beta]^n)$
        \State\algcmt{Compute the range of the uniform distribution}
        \State $\ell\leftarrow p(\mathbf{x})-\alpha$ 
        \State $r\leftarrow \ell+\beta$
        \State\algcmt{Initialize the set of similar topologies}
        \State $\text{T}\leftarrow \emptyset$
        \For {each graph $G\in \mathcal{D}$}
            \State\algcmt{Transform $G$ into a directed graph $G'$}
            \State $G'\leftarrow \textsc{InduceOrientation}(G)$ 
            \State \algcmt{Apply importance sampling}
            \State $\mathbf{x}\leftarrow \textsc{ComputeFeatures}(G')$
            \State $p'\leftarrow p(\mathbf{x})$
            \State $\text{T}\leftarrow \text{T}\cup \begin{cases} \{G'\} & \text{w/ prob $\mathds{1}(\mathbf{x}\in [\ell,r])/p'$} \\ \emptyset & \text{otherwise}\end{cases}$
        \EndFor
        \State \Return \text{T}
        \EndFunction
    \end{algorithmic}  
\end{algorithm}

}
\noindent\textbf{Operator Population.} After generating graph topologies, \codename assigns a DL operator at each node in the generated graph. A DL operator describes the computations that will be applied to tensor given as input (e.g., matrix multiplication, convolution, activation). The DL operators assigned to nodes need to be (i) syntactically correct, i.e., they need to have correct configurations (e.g., the number of input and output arguments, the tensor dimensions) that are consistent with specifications of DL operators, and (ii) semantically consistent, i.e., the sequence (order) of operators within the generated graph needs to resemble realistic DL operator sequences.

To ensure \emph{syntactic correctness}, we can convert this problem into one of \emph{constraint satisfaction}, i.e., given a set of constraints we need to find a solution that satisfies them. In our context, given a set of syntactic constraints for DL operators, we need to find an assignment of DL operators to the nodes of the graph, which satisfies the given syntactic constraints.  We use Z3~\cite{z3}, an SMT solver to produce the assignment of the operators to the sentinel nodes.  

Z3 takes as inputs (i) the graph topology, (ii) the list of operators, and (iii) their syntactic constraints, searches the solution space and returns a syntactically correct assignment of the operators to the nodes of the graph. For convolution and pooling, along with the operator we also need to specify the operator's kernel shape and number of input/output channels (for 2D convolutions). %

To ensure \emph{semantic consistency}, we need to quantify the \emph{likelihood} of an operator assignment. If the likelihood is high, the operator assignment is more likely to be semantically similar to real-world DL operator assignments. To compute the likelihood, we calculate probabilities of operator sequences generated by traversing the graph.  
   
\begin{algorithm}
    \begin{algorithmic}[1]
        \Function{AssignOperators}{$G, \text{pct}, \text{max\_solns}$}
        \State \text{Rules}$\ \leftarrow\textsc{GenerateRuleSet}(G)$
        \State \text{Solver}$\ \leftarrow \textsc{Z3Solver}$
        \State \text{S}$\ \leftarrow \emptyset$
        \State \algcmt{Loop until no new solutions or maximum reached}
        \While {$\operatorname{satisfiable}(S)$ \textbf{and} $|\text{Solns}|\leq \text{max\_solns}$}
            \State $S\leftarrow\textsc{GetSolution}(\text{Solver}, \text{Rules})$
            \State \algcmt{Find logprob for solution $S$}
            \State $p\leftarrow \operatorname{logprob}(S)$ 
            \State $\text{Solns}\leftarrow \text{Solns} \cup \{(S, p)\}$
            \State \algcmt{Prevent $S$ from being returned again}
            \State $\text{Rules}\leftarrow \text{Rules} \land (\neg S)$ 
        \EndWhile
        \State \Return $\textsc{TopPercentile}(\text{Solns}, \text{pct})$
        \EndFunction
    \end{algorithmic}
    \caption{Generating opcode specifications}
    \label{alg:specify_opcodes}
\end{algorithm}

Algorithm~\ref{alg:specify_opcodes} describes the operator assignment process. The \textsc{AssignOperators} procedure takes as input the graph topology and returns a set of operator assignments. Specifically, we repeatedly query the solver to find syntactically valid operator assignments. For each solution, we compute its likelihood, and record it (line 8). We also exclude the solution from being returned in a subsequent iteration. We repeat this procedure until either the solver claims that there are no other solutions (i.e. unsatisfiable) or when the number of solutions exceeds a predefined limit (line 5). We return the operator assignments that are both \emph{syntactically valid} and \emph{semantically likely}.

\noindent\textbf{Minor Modifications over Popular Models.} To handle the scenarios where the original protected model is structurally very similar to commonly-used popular DL models, e.g., the protected model is a ResNet-like model, \codename also generates graph topologies by modifying the topologies of popular DL models. Specifically, \codename generates new DNN-like graph topologies by adding and/or removing nodes in the existing graph topologies of popular DL models. Then, \codename fills DL operators to the newly added nodes using the process described above. In these cases, the opcodes of unperturbed nodes, except for the ones that are immediately adjacent to the perturbed nodes, are preserved.

\subsection{Optimization} \label{sec:detailed_optim}
After obfuscation, the set of $n(k+1)$ obfuscated subgraphs (including the original and generated sentinels) are given to the optimizer party. The optimizer applies graph transformations to each of the provided subgraphs to minimize their runtime behavior and provide performance speedups. 

The optimization step is carried \emph{independently} by the optimizer party on the obfuscated subgraphs. Note that \codename is largely \emph{agnostic} to the optimizer, since it does not make \emph{any} assumptions about the the optimizer's implementation other than that it preserves functional correctness. The optimizer will then return an optimized version of each obfuscated subgraph. 

\vspace{-8pt}
\subsection{De-obfuscation} \label{sec:detailed_deobfs}
\vspace{-4pt}
Upon receiving the optimized subgraphs from the optimizer, \codename \textit{reconstructs} the original model in its optimized form. It does so by extracting and concatenating the optimized ``real'' subgraphs. 

Assuming that the optimization procedure performed by the optimizer is functionally correct, the optimized subgraphs are functionally equivalent to the original subgraphs (up to numerical differences). Thus, when we reassemble the model using the optimized subgraphs, we obtain a computation that is defined by the composition of these subgraphs. If the subgraphs are functionally correct, their composition would also be functionally correct. In our implementation, the obfuscation step generates the optimized graph graph by connecting the input and output edges of each adjacent subgraph. This can be done using information about subgraph connections tracked when the graph was partitioned. Finally, the de-obfuscated graph is returned to the model owner as the optimized version of his original model. 

\vspace{-8pt}
\subsection{Parameterization}\label{sec:parameterization}
\vspace{-2pt}
\codename provides a number of tunable parameters to the model owner outlined in figure \ref{fig:parameters}. These parameters allow for tradeoffs between (a) the complexity of recovery by an adversary, (b) the computational overhead for the optimizer, and (c) the quality of model optimizations (in particular, the slowdown compared to optimizing without partitioning). 
\begin{figure}[h]
    \centering
    \begin{tabular}{c|p{0.35\textwidth}}
        \textbf{Name} & \textbf{Description} \\
        \hline
        $n$ & Number of \emph{graph partitions} generated from the protected graph \\
        $k$ & Number of \emph{sentinel subgraphs} generated per protected subgraph
    \end{tabular}
    \caption{List of tunable parameters provided by \codename}
    \label{fig:parameters}
\end{figure}

We tabulated the precise tradeoffs as a result of these parameters in Figure \ref{fig:tradeoff}. These tunable parameters allow the model owner to tradeoff some potential speedups for additional and stronger obfuscation. 

\textbf{Manual Optimization.} We note that while a $\mathcal{O}(k)$-fold increase is acceptable for an automatic optimizer or tensor compiler. However, if each subgraph requires manual engineering efforts to optimize, this overhead would be prohibitive, and \codename would be ineffective for these cases. However, we observe that most manual efforts are spent on development and tuning of the machine learning compiler instead of manually applying optimizations on each subgraph.
\section{Evaluation}
\vspace{-4pt}
\subsection{Methodology}
\vspace{-2pt}
\textbf{Runtime Environment.} \codename uses ONNX~\cite{onnx} for intermediate model representation, i.e., the initial DL model, its intermediate computational graph representation, and the optimized version of the given DL model are represented using the ONNX format. 

To demonstrate optimizer agnosticism, we use ONNXRuntime and Hidet for model optimizations and inference. ONNXRuntime is a performant optimizer and inference engine and Hidet~\cite{ding2023hidet} is a state of the art machine learning compiler.

We conduct our experiments on a \texttt{a2-highgpu-1g} instance on Google Cloud with 85GiB of RAM and an NVIDIA A100 GPU. 

\noindent\textbf{Models.} We evaluate \codename using representative widely-used convolutional neural networks (CNNs) that perform image classification as well as BERT-like language models (listed in figure~\ref{fig:adversary_cost}). We obtained their implementations through the \texttt{torchvision} package~\cite{torchvision} and HuggingFace Model Hub~\cite{huggingface}. 

\begin{figure}[h]
    \centering
    \begin{subfigure}{\columnwidth}
        \centering
        \includegraphics[width=0.9\columnwidth]{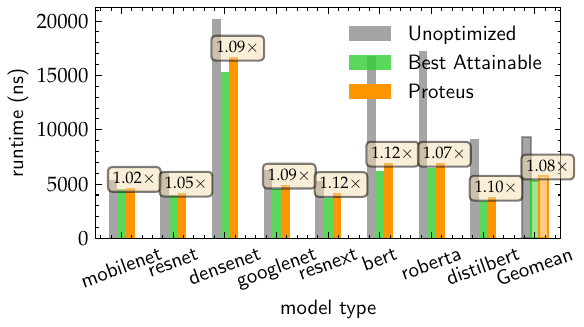}
        \caption{Evaluation with ONNXRuntime}
    \end{subfigure}
    \begin{subfigure}{\columnwidth}
        \centering
        \includegraphics[width=0.9\columnwidth]{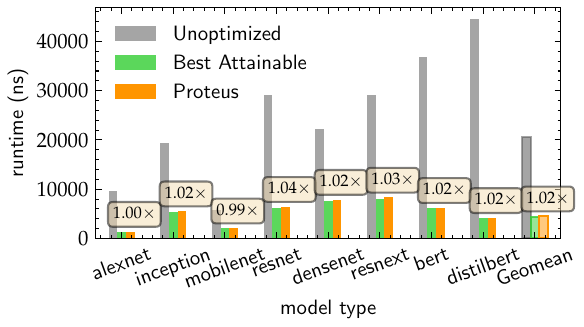}
        \caption{Evaluation with Hidet}
    \end{subfigure}
    \caption{Execution time of DL models achieved by all evaluated schemes. The slowdown of Proteus over Best Attainable is labelled above each model.}
    \label{fig:attainable_speedups}
\end{figure}

\vspace{-6pt}
\subsection{Performance Efficiency}\label{sec:eval-performance}
\vspace{-2pt}
We first explore the ability of \codename to retain the effectiveness of optimizations that generate performance speedups. ONNXRuntime performs a series of graph-level optimizations, from basic techniques such as constant folding to more complex operator fusion. 
Figure \ref{fig:attainable_speedups} depicts the resulting runtime for different DNNs, measured using the ONNXRuntime profiling tool across $500$ iterations and computing the geometric mean of a single iteration. 
We evaluate three different mechanisms: (i) \emph{Unoptimized}: without enabling \emph{any} graph-level optimizations in the baseline graph, (ii) \emph{Best Attainable}: enabling the best-performing graph-level optimizations available for the initial graph, and (iii) \emph{\codename}: using \codename to protect confidentiality of the model by partitioning into subgraphs, and then enabling the best-performing graph-level optimizations available for each subgraph.

We observe that on average \codename enables performance speedups close to the speedup of the optimizer without the confidentiality protection (within $8\%$ of the maximum speedup on average and at most $12\%$). The small loss in performance speedups is due to the partitioning approach which reduces the effectiveness of some optimization techniques. For example, if a \texttt{conv} operator is following by an \texttt{add} operator in the original graph but the two are partitioned into different subgraphs, then fusion cannot be done between them. Performance loss due to graph partitioning is also dependent on the choice of the optimizer, i.e., the graph-level substitutions enabled by each particular optimizer. However, we argue that since various tensor compilers typically perform local graph transformations, we see similar behaviours with Hidet \cite{ding2023hidet} and expect similar performance trends for other optimizers.

Loss of performance optimization opportunities is correlated with the average size of the subgraphs and this is further investigated in appendix \ref{sec:appendix_speedup_lost}. We provide subgraph size as a tunable parameter as larger subgraph sizes would require more sentinels and thus, higher overheads for optimization. A subgraph size of $8-16$ offers a sweet spot where performance loss is less than $10\%$ on average and incurs only small optimization overheads and is what we use in the remaining evaluations for \codename. 

The optimization overhead is also correlated with the number of sentinels generated per real partition. We include specifics of the tradeoff in appendix \ref{sec:appendix_parameterization}. 

\subsection{Protection of Confidentiality}\label{sec:eval-confidentiality}

We demonstrate that \codename generates sentinel subgraphs that are difficult to differentiate from real subgraphs by (i) evaluating graph statistics of sentinel subgraphs (Section~\ref{sec:statistics}), (ii) devising a learning-based adversarial attack on sentinel subgraphs (Section~\ref{sec:adversary-attack}), and (iii) evaluating the feasibility of manual and expert intervention (Section~\ref{sec:user-survey}).

\subsubsection{Statistical Quality of Sentinel Subgraphs}\label{sec:statistics}

\begin{figure}
    \centering
    \includegraphics[width=0.4\textwidth]{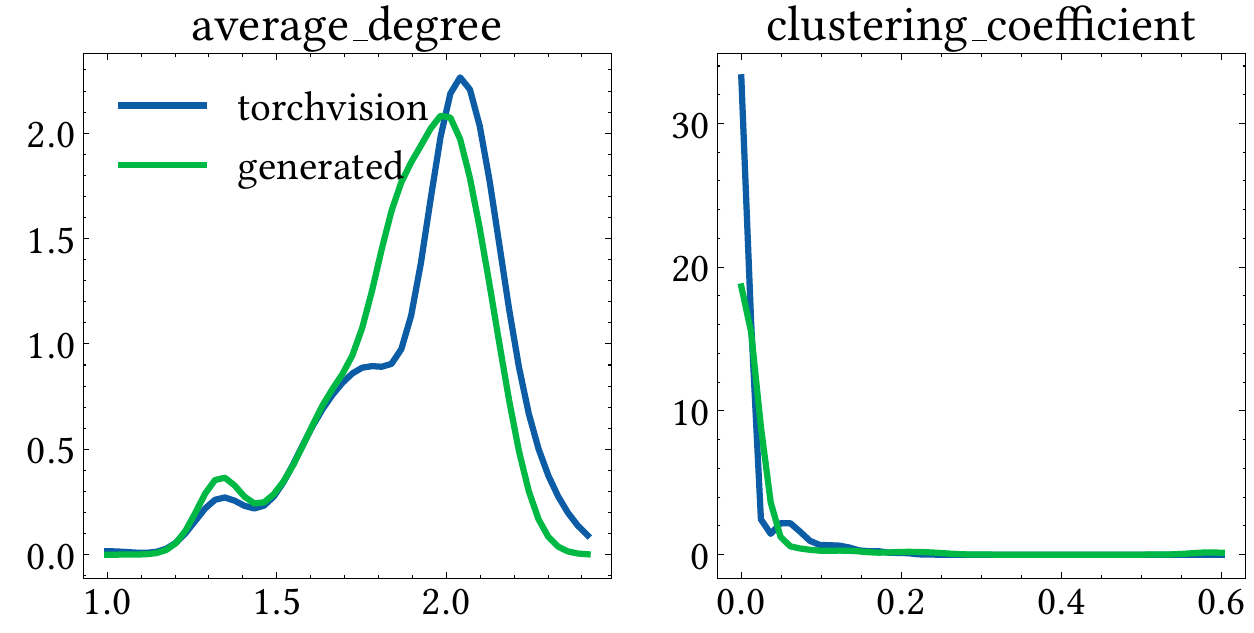}
    \caption{Comparing distributions of graph statistics between real and \codename-generated subgraphs}
    \label{fig:graph_metrics_partial}
    \vspace{-10pt}
\end{figure}

We evaluate the quality of generated sentinel graphs with \codename by comparing their distributions on various graph statistics with that of real graphs. Figure~\ref{fig:graph_metrics_partial} compares the distributions between real and \codename-generated graphs of their (a) average degree and (b) clustering coefficient.
Appendix~\ref{sec:appendix_graph_metrics_sentinels} evaluates two additional metrics. We observe that sentinel subgraphs of \codename have very similar distributions to that of real graphs in all evaluated metrics and would not distinguish sentinels from real graphs. We conclude that \codename is robust against mechanisms that may leverage graph heuristics or employ statistical distributions of various metrics to identify the protected graph.

We note that even when using fewer metrics in algorithm \ref{alg:sample_topologies}, we still observe distributional similarities in other graph metrics. This is due to GraphRNN's ability to learn complex edge dependencies by observing real topologies, then imitating it to generate realistic sentinel topologies and leads to robustness across metrics. 

\subsubsection{Learning-Based Adversarial Attack}\label{sec:adversary-attack}
\newcolumntype{a}{>{\columncolor{Gray}}l}
\newcolumntype{w}{>{\columncolor{White}}l}

\definecolor{proteuscolor}{RGB}{206, 255, 194}
\newcolumntype{k}{>{\columncolor{proteuscolor}}l}

\begin{figure*}[th!]
\scriptsize
\centering
\begin{tabular}{l|ll|wwa|wwk}
\multirow{2}{*}{Protected Model} & \multirow{2}{*}{$n$} & \multirow{2}{*}{$k$} & \multicolumn{3}{|w}{Random Opcodes} & \multicolumn{3}{w}{\codename (Ours)} \\
 & & & Specificity & $\min(\gamma)$ & Candidates & Specificity & $\min(\gamma)$ & Candidates \\
 \hline
densenet & 19 & 20 & 0.000 & 1.000 & \num{13248496640331000000000000} & 0.338 & 0.757 & \num{8329178044116460000000} \\
googlenet & 11 & 20 & 0.990 & 0.356 & \num{7} & 0.346 & 0.899 & \num{4301155180508} \\
inception & 19 & 20 & 0.970 & 0.784 & \num{7692} & 0.229 & 0.910 & \num{122698654435473000000000} \\
mnasnet & 11 & 20 & 1.000 & 0.019 & \num{1} & 0.117 & 0.944 & \num{95881428358270} \\
resnet & 10 & 20 & 1.000 & 0.100 & \num{1} & 0.451 & 0.908 & \num{61194460969} \\
 mobilenet & 11 & 20 & 0.607 & 1.000 & \num{26579839891} & 0.135 & 0.977 & \num{77240354626497} \\
bert & 16 & 20 & 0.996 & 0.474 & \num{3} & 0.910 & 0.653 & \num{13670158} \\
roberta & 16 & 20 & 0.990 & 0.634 & \num{20} & 0.862 & 0.799 & \num{1543329068} \\
xlm & 25 & 20 & 1.000 & 0.300 & \num{1} & 0.906 & 0.816 & \num{298934345718}
\end{tabular}

\caption{Search space reduction for learning-based adversary}
\label{fig:adversary_cost}
\vspace{15pt}
\end{figure*}

Given the mechanism proposed in this work, adversaries with the objective to recover the original protected graph fundamentally need to decide if a particular subgraph in the bucket is a \codename-generated sentinel graph or part of the original protected graph. 

In this section, we put ourselves into the position of one such adversary who attempts this task with a learning-based approach. Particularly, we evaluate the effectiveness of using a graph neural network (GNN) to perform such differentiation and investigate if the classifier helps to reduce the search space to one that compromises the model architecture. 

\noindent\textbf{Classifier Architecture.} The classifier network accepts a graph as an input and outputs its confidence that the given graph is a sentinel. We depict the architecture of the classifier in Figure \ref{fig:gnn_architecture}, and we elaborate further in Appendix \ref{sec:appendix_classifier_architecture}.

\begin{figure}
    \centering
    \includegraphics[width=0.3\textwidth]{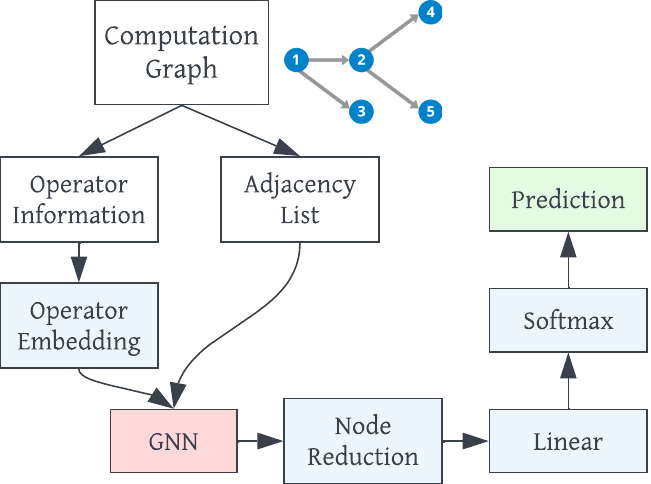}
    \caption{Architecture of GNN classifier}
    \label{fig:gnn_architecture}
    \vspace{-15pt}
\end{figure}

\noindent\textbf{Datasets.} During the training and evaluation of this learning-based adversary, we task the GNN-based adversary to differentiate real model subgraphs with the following:
\begin{enumerate}[leftmargin=*,noitemsep,topsep=0pt]
    \item \emph{Random opcodes on \codename-generated topologies.} We use the GraphRNN topologies with random operators. 
    \item \emph{Sentinel graphs from \codename.} We run the entire pipeline, using both GraphRNN and Z3. 
\end{enumerate}
In our experiments, we task \codename with protecting one model at a time. To do so, we test the adversary on the protected model after training the classifier model on the remaining models. 

\noindent\textbf{Attack Mechanism.} 
The GNN classifier outputs a confidence $y\in [0,1]$ of the graph being sentinel. The adversary would fix a decision boundary $\gamma$ such that a graph is eliminated as fake when $y\geq \gamma$. We make a pessimistic assumption that the adversary obtains $\gamma$. We note that the adversary must not erroneously eliminate any real subgraphs.

\noindent\textbf{Metrics.} We can measure the \emph{sensitivity} (how many real subgraphs are correctly classified) and its \emph{specificity} (how many fake subgraphs are correctly classified) of the adversary. Specifically, for each of the protected models we test and our choice of $n$, $k$ and $\gamma$, we can measure the classifier's sensitivity (denoted $\alpha$) and specificity (denoted $\beta$).

As established earlier, we must have $\alpha=1$ such that all \emph{real} subgraphs are correctly classified. In this case, for each of the $n$ real subgraphs, approximately $(1-\beta)k$ of its sentinel graphs are misidentified, resulting in a total number of $1+(1-\beta)k$ candidates for this particular subgraph. This results in a search space with size:
\[
    [1+(1-\beta)k]^n
\]
within which the protected graph is hidden. 

\noindent\textbf{Search Space Reduction.} In Figure \ref{fig:adversary_cost}, we compared the sizes of reduced search spaces with the learning-based approach. For each model type, we plotted the specificity ($\beta$) as well as the minimum decision threshold ($\gamma$) such that no real subgraphs are incorrectly identified. Using the cost computed above, we also tabulated the number of candidates in the reduced search space. The above is done for both the baseline (random opcode population) and \codename. 

We find that the resulting search space for differentiating \codename-generated graphs from real graphs is orders-of-magnitude larger than that of the baseline (random opcodes). In many cases, a single candidate remains for the baseline, therefore making recovery trivial. Thus, such attacks are effective when the sentinels are not appropriately generated, as addressed by \codename.

\subsubsection{User Survey on Sentinel Graphs}\label{sec:user-survey}

To evaluate the realism of the graphs and the possibility of manually intervention by experts to identify sentinels, we conducted a survey amongst researchers in ML (details of the methodology are in Appendix~\ref{sec:appendix-survey}). The survey contains 20 computational graphs and asks the participants to classify them as either real (i.e., taken from a real popular ML model) or fake (i.e., generated by \codename). Out of 13 participants, the average accuracy is $52\%$, similar to that of random guesses. The survey can be accessed \href{https://tiny.cc/zswcvz}{here} and contains randomly chosen sample sentinel graphs that demonstrate the infeasibility of distinguishing sentinels simply by visual inspection from experts.
\vspace{-6pt}
\section{Case Studies}
We present two archetypal scenarios of graph-level model optimization: in the first case, the proposed model is unique and dissimilar to existing ones; and in the second, the proposed model largely resembles one that is widely used. In appendix \ref{sec:appendix_case_study}, we also evaluate the potential of GNN adversary presented in Section~\ref{sec:adversary-attack} in these case studies and present some visual examples of misclassified graphs.

\subsection{Optimizing NAS Model}
In the first scenario, the user optimizes a more ``exotic" model that is very dissimilar to existing ones. For this purpose, we sample a model from NATSBench's~\cite{natsbench} search space. Optimizing this model with ONNXRuntime results in a \emph{slowdown} of $2.15\times$. This is potentially due to some optimizations that are typically beneficial but turned out harmful for this particular exotic model. When this model is optimized with \codename, a similar outcome (slowdown of $2.164\times$) can be observed. Furthermore, the search space size is $1.18\times 10^{21}$ for the GNN adversary.

\subsection{Optimizing a ResNet-like Model}
In the second case study, the user attempts to optimize a model that resembles ResNet, SEResNet \cite{hu2019squeezeandexcitation}. The main difference lies in the addition of squeeze-excitation blocks. 
In the ideal case of optimizing the model directly, the maximum attainable speedup is $1.663\times$. With \codename, we obtain a speedup of $1.494\times$, representing a $\approx 10\%$ penalty. Furthermore, the search space size is $1.22\times 10^{87}$ for the GNN adversary.

\vspace{-6pt}
\section{Conclusion}
In this paper, we introduce and motivate a novel and unexplored research question of confidential compiler graph-level optimization for deep learning models. Our proposed solution, \codename, obfuscates the original model within realistic artificially generated computational graphs. \codename is largely agnostic to the optimization approach and is thus generally applicable. The incurred computational overhead is trivial when using modern graph-level compilers. 
We demonstrate \codename's robustness against learning-based, heuristic-based, and manual approaches, that are unable to distinguish the sentinel and protected subgraphs. A limitation of \codename is that the additional sentinels make manual intervention more expensive (but still not infeasible when $k=20$). The graph partitioning used in \codename could also undermine optimization opportunities. Reducing the number of sentinels and partitioning more effectively are areas for future exploration. We hope that our work provides a first step for future work on confidential compiler optimization in deep learning.

\section{Artifact Instructions}
Our artifact provides the code to reproduce two of our main results - those in figure \ref{fig:attainable_speedups} demonstrating the partition-optimize-reassemble routine preserves the optimization speedups and figure \ref{fig:adversary_cost} showing the difficulty of graph recovery by a learning based adversary. 

The \codename source code and scripts are available at \texttt{github.com/proteus-mlsys24/mlsys24-artifact}. 

For the full artifact appendix, please refer to appendix \ref{sec:appendix_artifact}. 

\newpage
\bibliography{ref}
\bibliographystyle{mlsys2024}

\appendix
\section{Appendix}
\subsection{GraphRNN Induce Orientation}\label{sec:appendix_induce_orientation}
The function \textsc{InduceOrientation} takes as input a graph $G=(V,E)$ and returns a directed graph $G'=(V,E')$, replacing undirected edges of $E$ with directed edges in $E'$. In line 2, we first find the endpoints $u,v\in V$ of the diameter of the graph. Next, we traverse $G$ with BFS beginning from the endpoint node $u$ (line 3), and record the order in which nodes are accessed in the graph traversal, which is stored in the array $ord$. Using $ord$, we determine the orientation of each edge $e\in E$: we orient the edges to point from the node with smaller $ord$ to the one with greater $ord$. The directed edges are collected in $E'$, and the resulting acyclic directed graph $G'=(V,E')$ is returned.

\begin{algorithm}[h]
    \begin{algorithmic}[1]
        \Function{InduceOrientation}{$G=(V,E)$}
            \State \algcmt{Find diameter endpoints}
            \State $u,v\leftarrow \textsc{Diameter}(G)$ 
            \State \algcmt{Record BFS traversal order}
            \State $ord[] \leftarrow \textsc{BFS}(G,u)$ 
            \State \algcmt{Set of directed edges}
            \State $E'\leftarrow\emptyset$ 
            \For{$e\coloneqq (u',v') \in E$}
                \If{$ord[u']<ord[v']$}
                \State Add $u'\to v'$ to $E'$
                \Else
                \State Add $v'\to u'$ to $E'$
                \EndIf
            \EndFor
            \State \algcmt{$G'$ is an acyclic orientation}
            \State \Return {$G'\coloneqq (V,E')$} 
        \EndFunction
    \end{algorithmic}
    \caption{Induce Orientation on GraphRNN Graphs}
    \label{alg:induce_orientation}
\end{algorithm}

\subsection{Parameterization}\label{sec:appendix_parameterization}
\codename provides a number of tunable parameters to the model owner outlined in figure \ref{fig:parameters}. These parameters allow for tradeoffs between (a) the complexity of recovery by an adversary, (b) the computational overhead for the optimizer, and (c) the quality of model optimizations (in particular, the slowdown compared to optimizing without partitioning). 
\begin{figure}[h]
    \centering
    \begin{tabular}{c|p{0.35\textwidth}}
        \textbf{Name} & \textbf{Description} \\
        \hline
        $n$ & Number of \emph{graph partitions} generated from the protected graph \\
        $k$ & Number of \emph{sentinel subgraphs} generated per protected subgraph
    \end{tabular}
    \caption{List of tunable parameters provided by \codename}
    \label{fig:parameters}
\end{figure}

We tabulated the precise tradeoffs as a result of these parameters in Figure \ref{fig:tradeoff}. These tunable parameters allow the model owner to tradeoff some potential speedups for additional and stronger obfuscation. 

\begin{figure}[h!]
    \centering
    \begin{tabular}{p{15em}|p{6em}}
        \textbf{Item} & \textbf{Cost} \\
        \hline
        Recovery cost of adversary & $\mathcal{O}((k+1)^n)$ \\
        Computational overhead of optimizer & $\mathcal{O}(k)$ \\
        Quality of model optimizations & See figure \ref{fig:partition_speedups}
    \end{tabular}
    \caption{Tradeoffs resulted from \codename's tunable parameters}
    \label{fig:tradeoff}
\end{figure}

For the DNNs in our evaluation, optimizing the original model takes 6s on average and up to 22s with Hidet on an AMD EPYC 7282 CPU. If Proteus produces 50 sentinel graphs, thus optimizing both the original as well as sentinels would take 5 minutes on average and up to 18 minutes. \codename results in a $k$-fold increase in compilation time, where $k$ is the number of sentinels generated per partition, and as demonstrated above this cost is not prohibitive. 

\subsection{Subgraph Size vs. Slowdown}\label{sec:appendix_speedup_lost}
As suggested in section \ref{sec:eval-performance}, many optimizations would be ineligible due to the small graph size (as optimizations cannot be applied \emph{across} partitions). \codename provides the number of subgraphs $n$ as a tunable parameter to the user. In this section we evaluate the correlation between subgraph size and slowdown compared to optimal (where the entire graph is optimized as a whole). 

\begin{figure}[h]
    \centering
    \includegraphics[width=0.4\textwidth]{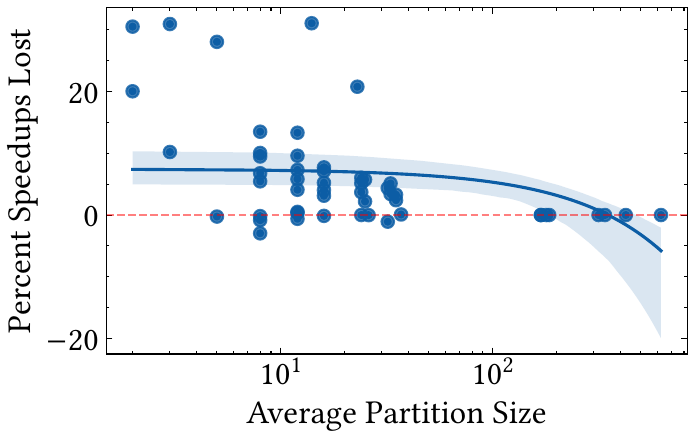}
    \caption{Average Subgraph Size vs. \% Performance Loss with \codename}
    \label{fig:partition_speedups}
\end{figure}

Figure~\ref{fig:partition_speedups} evaluates the trade-off between the size of the subgraphs and potential performance loss across all our evaluated DL models. We normalize performance loss as percentage over \emph{Best Attainable}. Each point in the figure represents the inference latency for a model and one of the three setups above. 
We observe that with very small subgraph sizes, \codename incurs small performance losses. However, when the average subgraph size becomes large, \codename achieves negligible performance losses.

\subsection{Graph Metrics of Generated Sentinels}\label{sec:appendix_graph_metrics_sentinels}
Let us consider $\mathcal{D}$ to be the distribution of generated sentinel graphs, and let $G$ be the real subgraph. The distribution of various graph characteristics for $\mathcal{D}\cup\{G\}$ should not be significantly different from the original distribution $\mathcal{D}$. Otherwise, an adversary could distinguish the protected subgraph by learning the characteristics of the generated sentinels. For this, it is necessary that the sentinel subgraphs are realistic and similar to real world subgraphs.

\begin{figure}[h]
    \centering
    \includegraphics[width=0.45\textwidth]{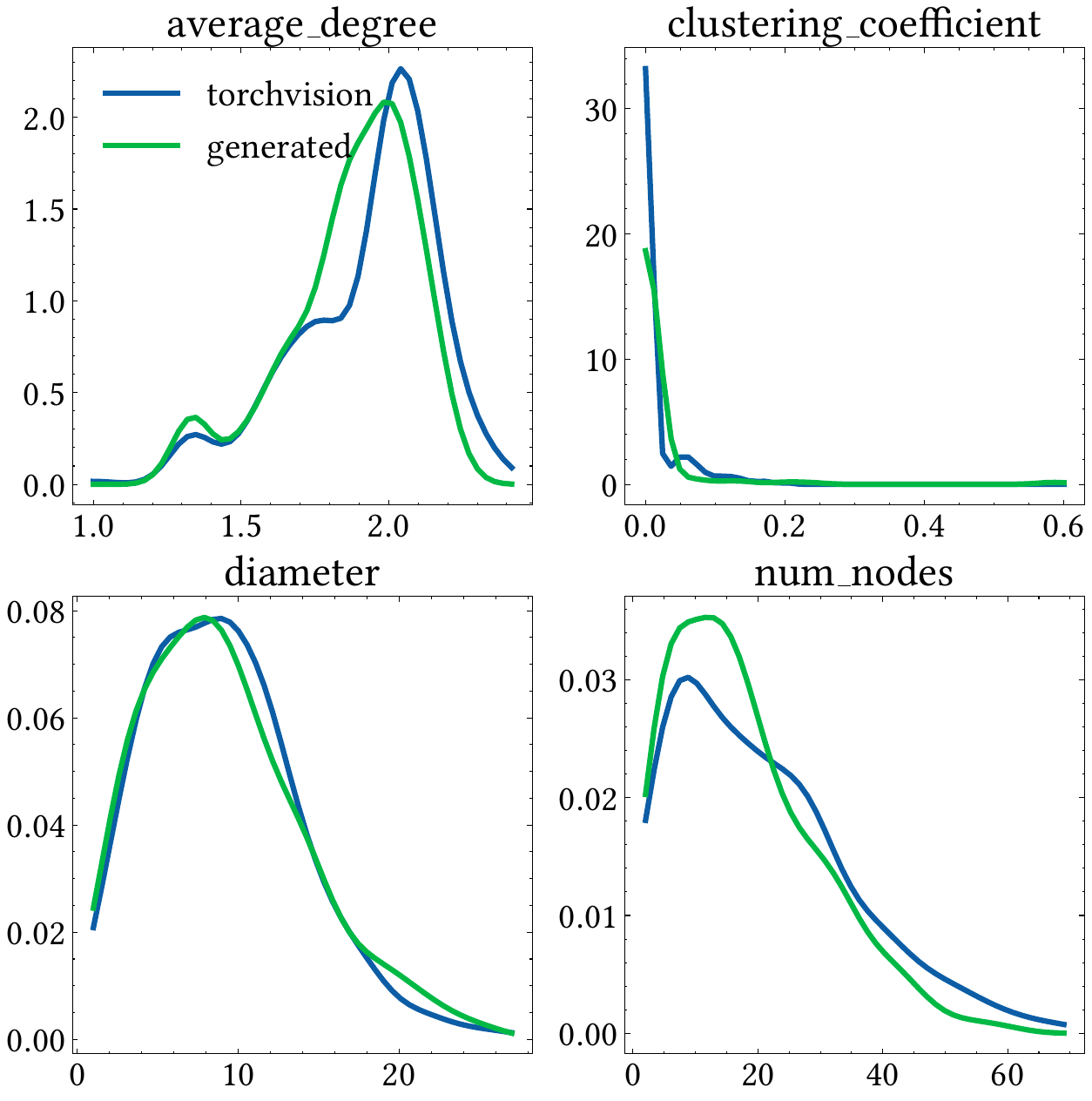}
    \caption{Comparing distributions of graph statistics between real and \codename-generated subgraphs}
    \label{fig:compare_graph_stats}
\end{figure}
Figure \ref{fig:compare_graph_stats} presents the distributions of (i) the real-world subgraphs extracted from \texttt{torchvision} models, and (ii) the sentinel (artificially) generated subgraphs produced by \codename  (algorithm \ref{alg:sample_topologies}) using different graph statistics. We use the set of graph statistics explained in Section \ref{sec:detailed_sentinel_gen}, i.e., (a) average degree, (b) clustering coefficient, (c) graph diameter and (d) number of nodes.

The X axis represents the value for the particular metric and the Y axis represents the probability density. From the figures we observe very little statistical difference between the two groups. Thus, we conclude that the sentinel subgraphs produced using our approach are highly realistic, resembling real world subgraphs and it would be very difficult for a potential adversary making use of these graph statistics from to differentiate real graphs from the sentinels.

\subsection{Classifier Architecture}\label{sec:appendix_classifier_architecture}
The classification network performs graph convolutions using SAGEConv \cite{graphsage} with the objective to learn features of the local neighborhood for each node in the graph. After that, aggregation is done across the nodes with mean reduction. This step essentially generates a hidden representation for the entire graph from the node representations after the GNN. The final classification is generated with linear layers acting on the reduced graph representation. 

\subsection{Adversary Cost}\label{sec:appendix_adversary_cost}
The adversary attempts to shrink its search space by eliminating fake graphs from the obfuscated bucket so as to narrow down on the ``possibly-real'' graphs. We especially note that it must not classify any real subgraphs as being sentinels, as that would eliminate the actual protected graph from its search space. Therefore it must fix some decision boundary $\gamma$, such that a graph is eliminated as fake when the classifier's confidence exceeds $\gamma$. 

The value of $\gamma$ defines how \emph{conservative} the adversary is in eliminating potential sentinel graphs. Decreasing $\gamma$ would reduce the number of ``potentially-real'' graphs, resulting in a smaller search space. However, by doing so the adversary also risks incorrectly eliminating a real subgraph. 

In practice, it would be difficult for an actual adversary to obtain the minimum value of $\gamma$ without a priori knowledge of the protected graphs, since it would not have access to the confidences of the classifier on them. However, we wish to establish an approximate \emph{lower bound} for the cost of attack by assuming the pessimistic case where the adversary obtains the optimal (i.e. lowest possible) $\gamma$ through some (perhaps statistical) means.

\subsection{Case Studies}\label{sec:appendix_case_study}
For both cases, we use our standard configuration, setting $n=\lfloor N/8\rfloor$ (where $N$ is the total size of the original model, such that subgraphs on average have $8$ nodes) and $k=50$. 

\subsubsection{Optimizing a NAS model}
\begin{figure}[h]
    \centering
    \begin{subfigure}{0.48\textwidth}
        \centering
        \includegraphics[width=0.5\textwidth]{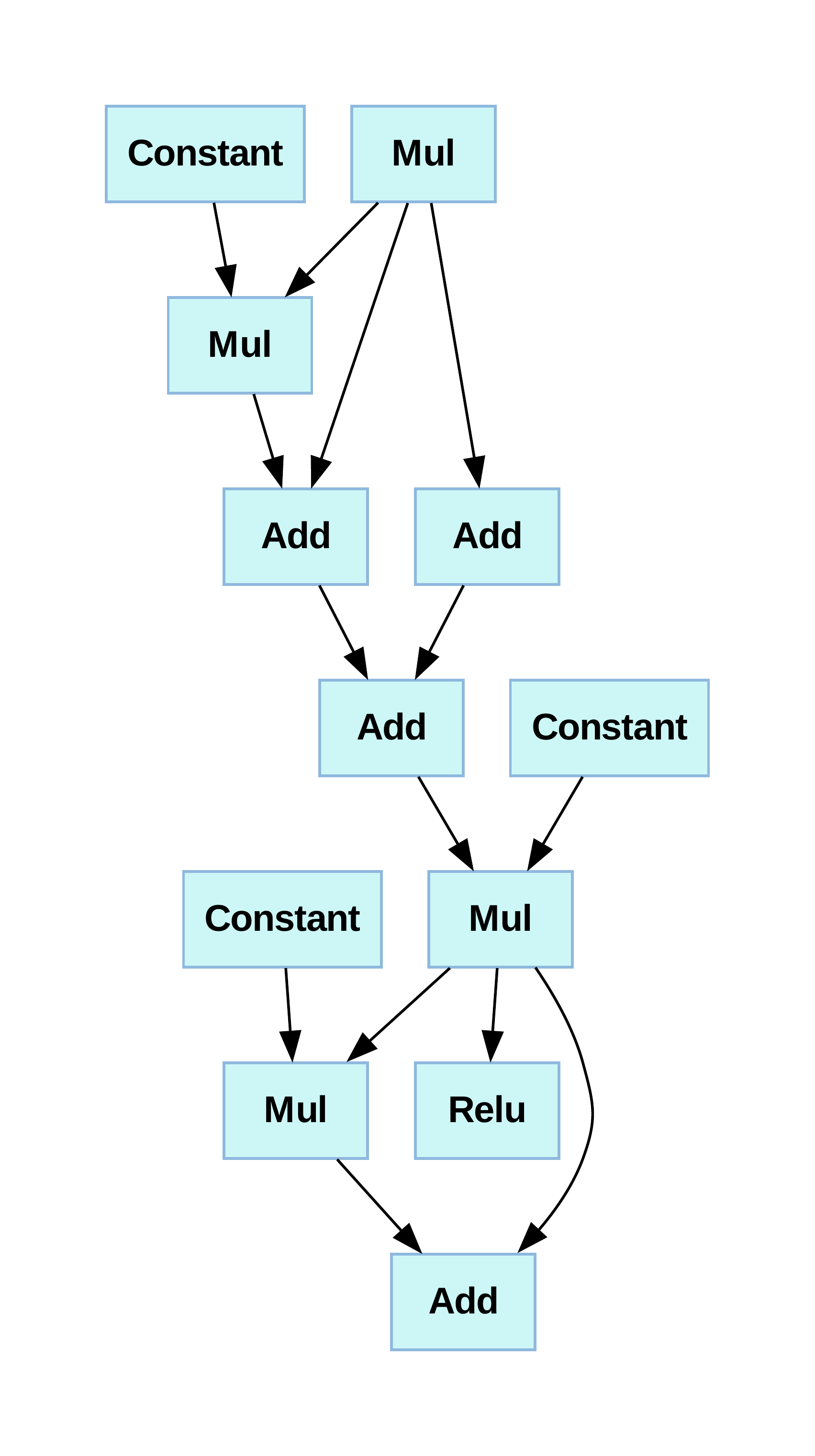}
        \caption{NATSBench: Sentinel incorrectly classified as real}
        \label{fig:subfiga}
    \end{subfigure}
    \begin{subfigure}{0.48\textwidth}
        \centering
        \includegraphics[width=0.5\textwidth]{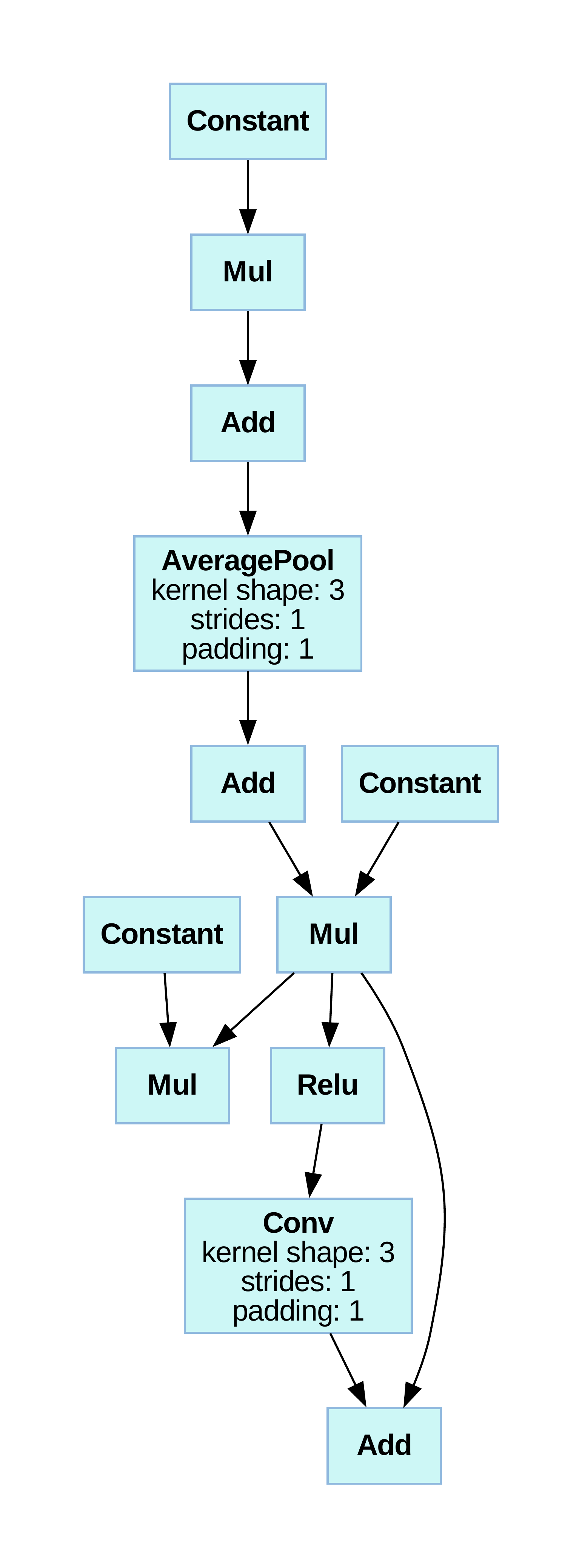}
        \caption{NATSBench: Real incorrectly classified as sentinel}
        \label{fig:subfigb}
    \end{subfigure}
    \caption{NATSBench Examples}
    \label{fig:case_study_natsbench}
\end{figure}

In addition to faithfully recreating the effects of optimizing the model directly, the obfuscation mechanism also is effective in protecting the model against recovery. In this case, the GNN classified many real graphs as being fake (in many cases concluding which with a $99\%$ certainty). Shown in Figure \ref{fig:case_study_natsbench} are two examples incorrectly classified subgraphs from NATSBench. 

Setting $\gamma$ accordingly resulted in a sensitivity of $84.9\%$. With $n=24$ and $k=50$, this resulted in $[50(1-0.849)]^{24}\approx 1.18\times 10^{21}$ candidates which the adversary would need to evaluate. 

\subsubsection{Optimizing a ResNet-like Model}
Due to the similarity between SEResNet with ResNet, we expect most of the real subgraphs to be classified correctly, but some fake graphs are classified as real. Setting $\gamma=0.79$, the sensitivity of the classifier is $44\%$. Using $n=83$ and $k=20$, the number of potential candidates is $[20(1-0.44)]^{83}\approx 1.22\times 10^{87}$.

\begin{figure}[h]
    \centering
    \begin{subfigure}{0.48\textwidth}
        \centering
        \includegraphics[width=0.5\textwidth]{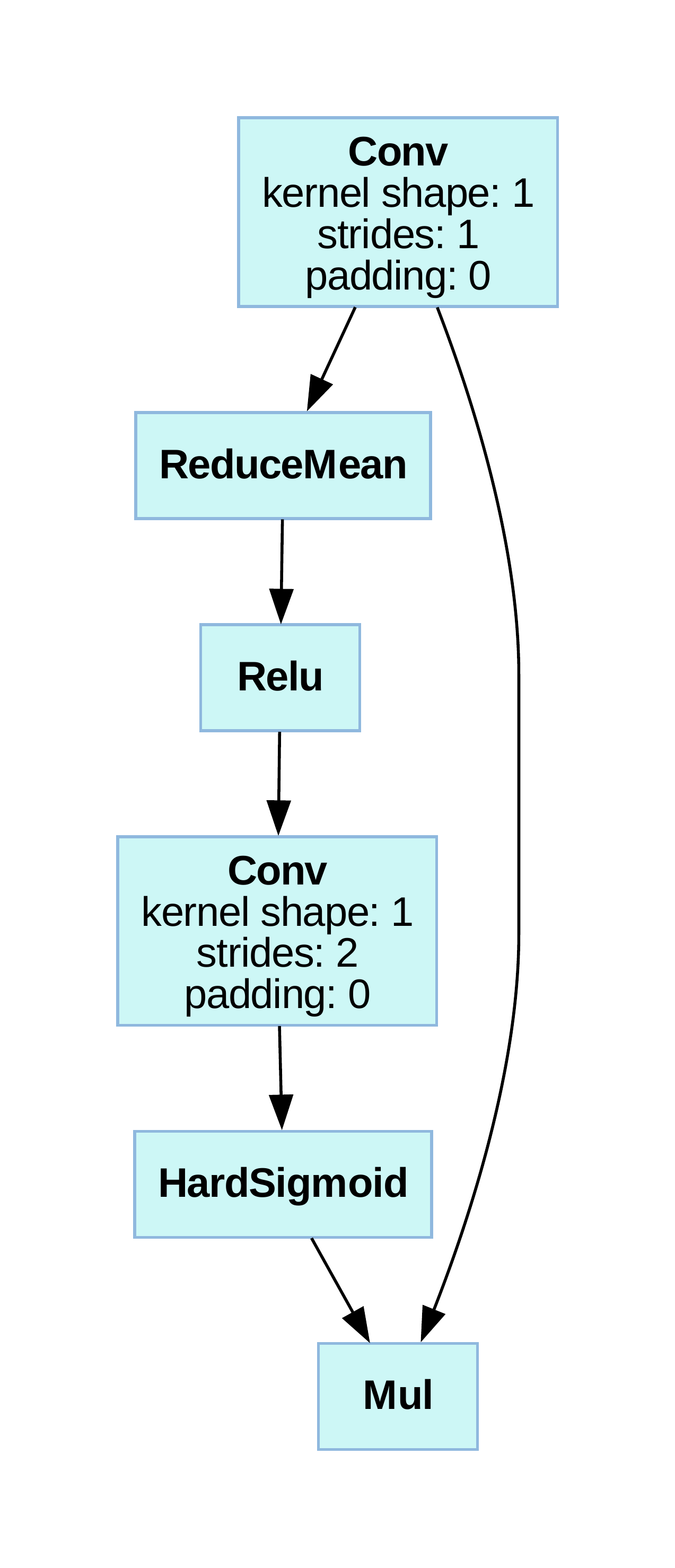}
        \caption{SEResNet: Sentinel incorrectly classified as real}
        \label{fig:subfiga}
    \end{subfigure}
    \begin{subfigure}{0.48\textwidth}
        \centering
        \includegraphics[width=0.5\textwidth]{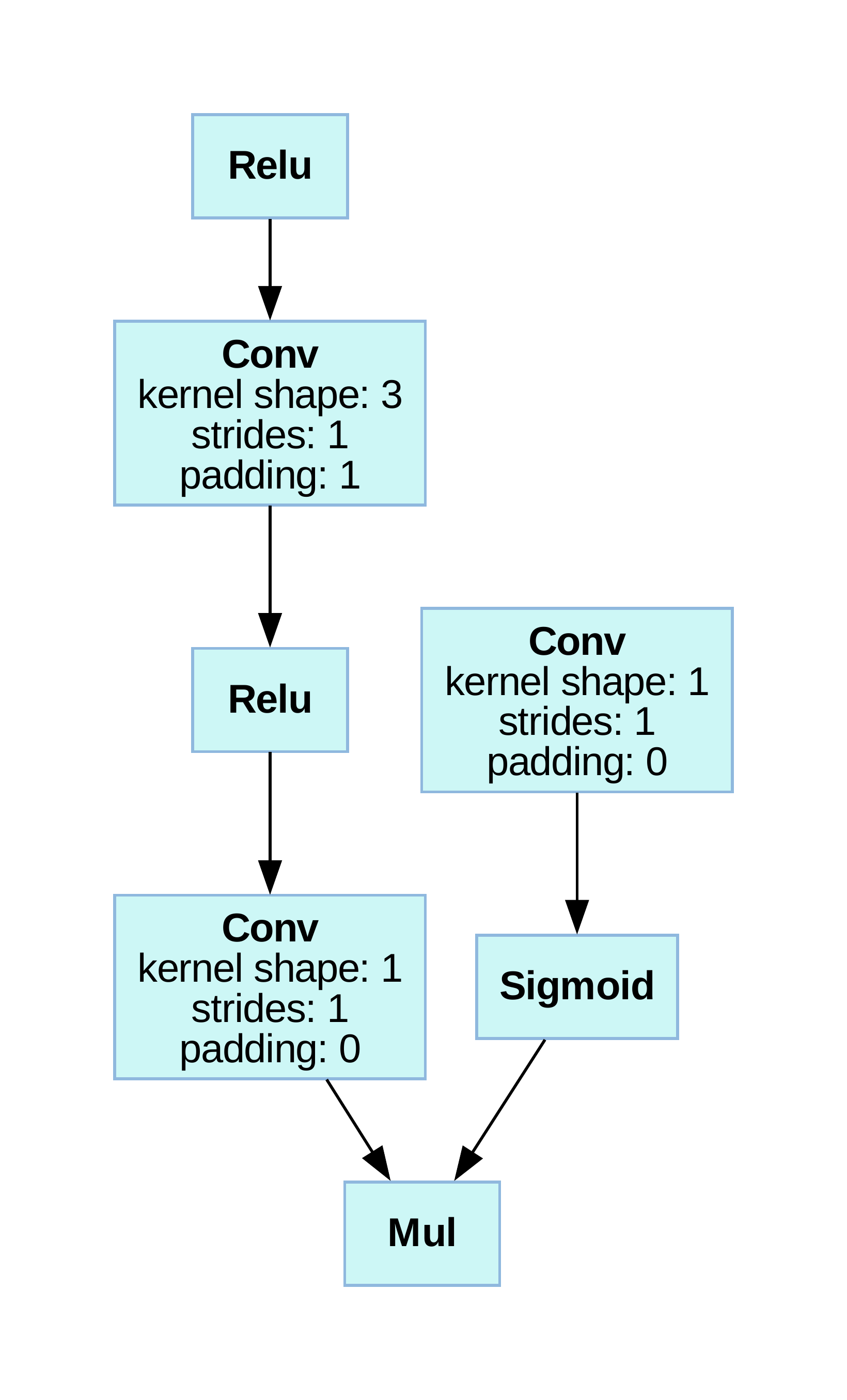}
        \caption{SEResNet: Real incorrectly classified as sentinel}
        \label{fig:subfigb}
    \end{subfigure}
    \caption{SEResNet Examples}
    \label{fig:case_study_seresnet}
\end{figure}

We show a few examples of incorrectly classified SEResNet examples in figure \ref{fig:case_study_seresnet}.

\subsection{Survey Study}\label{sec:appendix-survey}
We evaluate the possibility of manual identification by  experts to eliminate sentinels produced by \codename. To do so, we conducted an internel survey amongst ML researchers from the authors' primary institute. The survey contains 20 subgraphs consisting of 10 real subgraphs extracted models taken from \texttt{torchvision} and HuggingFace, and also another 10 subgraphs, which are \codename-created sentinels using the aforementioned real models. 

To generate graphs for the survey, \codename is configured to partition reals graphs into subgraphs of sizes $8-16$ nodes. Next, we use \codename to generate fake graphs from each of the real partitions. This way we create a pool of real subgraphs and another pool of \codename sentinel subgraphs. The 20 graphs are selected from the aforementioned pools at random with some small graphs filtered out.

For each of the 20 graphs, we ask the participants to select between two options: (i) real and (ii) fake subgraph. 

Out of 13 participants, the average accuracy is $52\%$, which demonstrates that the average guess of a participant is the same effective as guessing randomly. We conclude that the sentinel graph generation is robust against identification using expert knowledge and manual intervention through visual inspection. The survey can be accessed \href{https://tiny.cc/zswcvz}{here} and provides visual examples of sentinel and real graphs, demonstrating the realism of the sentinel graphs.

\section{Artifact Instructions}
\label{sec:appendix_artifact}
\subsection{Abstract}

Our artifact provides the code to reproduce two of our main results, that (a) the partition-optimize-reassemble workflow preserves the efficacy of optimizations, and (b) our graph obfuscation pipeline makes it difficult for a learning-based adversary to distinguish real graphs from sentinel ones. 

\subsection{Artifact check-list (meta-information)}

{\small
\begin{itemize}[leftmargin=*,]
  \item {\bf Data set: } Deep learning models in ONNX format from open source libraries such as \texttt{torchvision}
  \item {\bf Hardware: } Machine with NVIDIA A100 GPU
  \item {\bf How much disk space required (approximately)?: } 100GB
  \item {\bf How much time is needed to prepare workflow (approximately)?: } $<1$ hour
  \item {\bf How much time is needed to complete experiments (approximately)?: } $<10$ hours (significantly less if fast path is taken)
  \item {\bf Publicly available?: } Yes
  \item {\bf Code licenses (if publicly available)?: } Apache-2.0
  \item {\bf Data licenses (if publicly available)?: } Apache-2.0
  \item {\bf Workflow framework used?: } PyTorch, ONNX, ONNXRuntime, Hidet
  \item {\bf Archived (provide DOI)?: } \texttt{10.5281/zenodo.10977142}
\end{itemize}

\subsection{Description}

\subsubsection{How delivered}
The source code and scripts are available in the GitHub repository:
\texttt{https://github.com/proteus-mlsys24/mlsys24-artifact}. 

You can also find the archival version hosted on Zenodo at \texttt{10.5281/zenodo.10977142}. 

\subsubsection{Hardware dependencies}
Requires an NVIDIA A100 GPU to run the runtime-related experiments. Many CPU cores ($\approx 32$) is recommended to run the adversary experiment (figure 5).

\subsubsection{Software dependencies}

Since our experiments run inside Docker, Docker with NVIDIA GPU support (through \texttt{nvidia-docker}) is required. 

\subsection{Installation}
\begin{enumerate}[leftmargin=*,topsep=0pt]
    \item \textit{Docker.} Install Docker by following instructions at \href{https://docs.docker.com/engine/install/}{https://docs.docker.com/engine/install/}.
    \item \textit{nvidia-docker. } To use Docker with NVIDIA GPUs, we need to install \texttt{nvidia-docker}. To do so, follow the instructions at \href{https://docs.nvidia.com/datacenter/cloud-native/container-toolkit/latest/install-guide.html}{https://docs.nvidia.com/datacenter/cloud-native/container-toolkit/latest/install-guide.html}.
    \item \textit{Artifact Code. } Clone the repository to obtain the source code:
    \begin{minted}[autogobble,fontsize=\tiny,frame=lines]{text}
        $ git clone https://github.com/proteus-mlsys24/mlsys24-artifact
    \end{minted}
\end{enumerate}

\subsection{Experiment workflow}
\begin{itemize}[leftmargin=*,topsep=0pt]
    \item \emph{Figure 3a: ONNXRuntime Speedup} 
    
    Navigate to \texttt{figures/fig3a-ort-speedup} and run \texttt{run.sh}. 
    \item \emph{Figure 3b: Hidet Speedup} 
    
    Navigate to \texttt{figures/fig3b-hidet-speedup} and run \texttt{run.sh}.
    \item \emph{Figure 5: GNN Classifier Adversary}

    Navigate to \texttt{figures/fig5-gnn-classifier} and follow the instructions in \texttt{README.md}. 
\end{itemize}
\subsection{Evaluation and expected result}
\begin{itemize}[leftmargin=*,topsep=0pt]
    \item \emph{Figure 3a/b: Speedups} 
    
    These experiments should generate a figure named \texttt{speedups.pdf} similar to those in the paper. 
    \item \emph{Figure 5: GNN Classifier Adversary}

    For each model $m$, the number of candidates should be large. We further expect the number of candidates for $m$ to be much larger than \texttt{$m$\_randop} (the baseline with random opcodes). 
\end{itemize}
\subsection{Experiment customization}
The \texttt{proteus} Python package is available as a standalone package for model obfuscation. 

%


\end{document}